\documentclass[manuscript]{IEEEtran}

\usepackage[utf8]{inputenc}
\usepackage[T1]{fontenc}
\usepackage[english]{babel}
\usepackage{amsmath, amssymb, amsfonts, mathtools}
\usepackage{braket}

\usepackage{graphicx}
\usepackage{float}
\usepackage{subcaption}
\usepackage{caption}
\usepackage{cuted}
\usepackage{color}
\usepackage{tikz}
\usetikzlibrary{arrows.meta, positioning}

\usepackage{booktabs}
\usepackage{multirow}
\usepackage{tabularx}
\usepackage{longtable}
\usepackage{array}
\usepackage{adjustbox}
\usepackage[table]{xcolor}
\usepackage{enumitem}
\usepackage{array}
\newcolumntype{L}[1]{>{\raggedright\arraybackslash}p{#1}}
\newcolumntype{C}[1]{>{\centering\arraybackslash}p{#1}}
\newcolumntype{R}[1]{>{\raggedleft\arraybackslash}p{#1}}

\usepackage{comment}
\usepackage{xcolor}
\definecolor{lightgray}{gray}{0.9}
\definecolor{headerblue}{RGB}{64, 128, 255}

\usepackage{siunitx}

\usepackage{cite}
\usepackage{url}
\usepackage{hyperref}
\hypersetup{
    colorlinks=true,
    linkcolor=black,
    citecolor=black,
    urlcolor=blue,
}
\usepackage{orcidlink}

\usepackage{hyperref}
\hypersetup{colorlinks=true,citecolor=black,linkcolor=black,urlcolor=black}
\usepackage[nameinlink,noabbrev]{cleveref}
\Crefname{figure}{Fig.}{Figs.}

\usepackage{algorithm}
\usepackage{algpseudocode}
\usepackage{algorithmicx}

\newcounter{algorithmicinline}

\usepackage{romannum}

\begin{document}

\title{Resource-Constrained Quantum Implementation and Analysis of Mini-AES Cipher}

\author{
    Syed Shahmir \orcidlink{0009-0001-1634-2713}\IEEEauthorrefmark{1},
    Ghulam Murtaza\IEEEauthorrefmark{3},
    Elmahdi Bentafat \orcidlink{0000-0001-9801-5582}\IEEEauthorrefmark{1},
    Mohamed Abdallah \orcidlink{0000-0002-3261-7588}\IEEEauthorrefmark{1},
    Ala Al-Fuqaha \orcidlink{0000-0002-0903-1204}\IEEEauthorrefmark{1}, 
    Saif Al-Kuwari \orcidlink{0000-0002-4402-7710}\IEEEauthorrefmark{2},
    and Tasawar Abbas \orcidlink{0000-0002-6048-7346}\IEEEauthorrefmark{4} \\
    \IEEEauthorblockA{\IEEEauthorrefmark{1} {\small College of Science and Engineering, Hamad Bin Khalifa University, Doha, Qatar}} \\
    \IEEEauthorblockA{\IEEEauthorrefmark{2}{\small Qatar Center for Quantum Computing, College of Science and Engineering, Hamad Bin Khalifa University, Doha, Qatar}} \\
     \IEEEauthorblockA{\IEEEauthorrefmark{3} {\small QuantumRonics lab, National Science and Technology Park, NUST University, Scholars Ave, H-12, Islamabad, Pakistan.}} \\
    \IEEEauthorblockA{\IEEEauthorrefmark{4}
    {\small Department of Physics, COMSATS University Islamabad, 45550 Islamabad, Pakistan}}
}
\maketitle

\begin{abstract}
Quantum search algorithms, such as Grover's algorithm, will soon pose a threat to current encryption standards as quantum computing advances toward practical implementation. These algorithms rely on quantum circuits that map cryptographic algorithms into appropriate Hilbert spaces, thus linking algorithmic inputs and outputs through quantum superposition. However, implementing and testing such circuits on existing quantum simulators, such as Qiskit, encounters significant memory and computational bottlenecks that limit research progress. To address these simulation constraints, we present a resource-optimized quantum circuit implementation of a complete two-round Mini-AES encryption algorithm. Our approach systematically converts the cipher's algebraic structure using algebraic normal form (ANF) equations, mapping XOR operations to CNOT gates and AND operations to Toffoli gates. To reduce memory overhead, we implement an ancilla-reuse strategy utilizing mid-circuit measurements and resets. Additionally, we introduce an automated procedure that generates Qiskit-compatible quantum circuits for arbitrary S-boxes and cryptographic primitives, streamlining the development of quantum cryptanalysis implementations. Our implementation achieves full encryption and decryption in a quantum circuit with modest gate requirements that can be achieved by the current or near-future state of the art quantum computers. In addition, we provide system utilization metrics to validate our resource-constrained implementation. This compact and pedagogically accessible implementation establishes a practical foundation for quantum cryptanalysis research and post-quantum cryptographic analysis. 
\end{abstract}
\begin{table*}[t]
\centering
\scriptsize
\caption{Comparison of resource estimates for AES implementations in quantum circuits and Grover's search from prior studies.}
\label{com}
\resizebox{\textwidth}{!}{%
\begin{tabular}{|L{1.7cm}|L{4cm}|L{3.5cm}|L{3.5cm}|L{4cm}|}
\hline   \rowcolor{lightgray}
\textbf{Ref} & \textbf{Description} & \textbf{Qubits Used} & \textbf{Grover Estimates} & \textbf{T-depth / T-counts} \\
\hline
\raggedright Grassl et al. 2016~\cite{grassl_applying_2016} & Implementation and Grover’s estimates with reversible AES circuits (Clifford+T) & 
For AES-128: 984 qubits \newline For AES-192: 1,112 qubits \newline For AES-256: 1,336 qubits & 
For AES-128: 2,953 qubits \newline For AES-192: 4,449 qubits \newline For AES-256: 6,681 qubits &
AES-128: $T_{depth} \approx 1.19\cdot$2$^{86}$ \newline AES-192: $T_{depth} \approx 1.81\cdot$2$^{118}$ \newline AES-256: $T_{depth} \approx 1.41\cdot$2$^{151}$ \\
\hline
\raggedright Almazrooie et al. 2018~\cite{almazrooie_quantum_2018} & Explicit reversible AES-128 circuit (Itoh–Tsujii inversion for SubBytes, zigzag architecture) & For AES-128: 928-930 qubits + 48/96 ancilla & N/A & 
192,832 CNOTs, 150,528 Toffoli, and 1,370 Pauli-X  \newline SubBytes depth: 676 per call
\\\hline
\raggedright Jaques et al. 2020~\cite{jaques_implementing_2020} & Grover oracles in Q\# with depth-optimized S-boxes (Boyar–Peralta), improved key expansion & 
For In-place:  \newline 
\hspace*{1em} AES-128: 1,785 qubits  \newline 
\hspace*{1em} AES-192: 2,105 qubits  \newline 
\hspace*{1em} AES-256: 2,425 qubits  \newline 
For Shallow:  \newline 
\hspace*{1em} AES-128: 2,937 qubits  \newline 
\hspace*{1em} AES-192: 3,513 qubits  \newline 
\hspace*{1em} AES-256: 4,089 qubits
& 
For In-place: \newline 
\hspace*{1em} AES-128: 1,665 qubits \newline 
\hspace*{1em} AES-256: 4,609 qubits \newline 
For Shallow: \newline 
\hspace*{1em} AES-128: 3,329 qubits \newline 
\hspace*{1em} AES-192: 3,513 qubits \newline 
\hspace*{1em} AES-256: 6,913 qubits \newline   
&
AES-128: 54,400, $T_{depth} \approx$120 \newline 
AES-192: 60,928, $T_{depth} \approx$120 \newline 
AES-256: 75,072, $T_{depth} \approx$126 
\\\hline
\raggedright Jang et al. 2021~\cite{jang_grover_2021} & Optimized Grover oracle for S-AES (16-bit block/key) with lookup-based S-box and efficient MixColumns & For S-AES: 32 qubits  & Multiply circuit cost with 2$\times$2$\times \frac{\pi}{4}\sqrt{2^{16}}$ &
144 CNOTs, 96 Toffoli, 16 CCX, and 35 X gates, $T_{depth} \approx 47$ 
\\\hline
\raggedright Wang et. al. 2022~\cite{wang_quantum_2022} & Improved AES-128 \& Sample-AES quantum circuits by optimized key expansion, modified SubBytes and use of affine transformation &  For AES-128: 656 qubits  \newline For S-AES: 48 qubits & N/A &  
AES-128: 18,040 Toffoli, 101,174 CNOTs, and 1,976 X gates  \newline 
Sample-AES: 168 Toffoli, 364 CNOTs, and 75 X gates
\\\hline
\raggedright Our Work &  Mini-AES in quantum circuits by logic base optimization of affine transformation and direct basis encoding of keys &   For Mini-AES: 28 qubits & 
Convert all operations to Toffoli gates and multiply with $\frac{\pi}{4}\sqrt{2^{16}}$ & 
CNOT = 198, Toffoli = 160, Reset = 156, NOT = 58, and SWAP = 26
\\\hline
\end{tabular}%
    }
\end{table*}
\section{Introduction}
In the context of cryptography, the quantum algorithms are usually divided into two main categories. The first category include the quantum Fourier transform-based algorithms, such as Shor’s algorithm for integer factorization and quantum phase estimation. The second category consists of search algorithms like Grover’s algorithm and its generalizations, which provide a provably optimal quadratic speedup applicable to a broad range of use cases. Recent developments in quantum computing have demonstrated that quantum algorithms such as Shor's and Grover's algorithms pose significant threats to the security of many cryptographic systems~\cite{zheng_implementation_2007, sakhi_grover_2011, hossain_faruk_review_2022}. Asymmetric encryption can theoretically break by Shor's algorithm by factoring large integers exponentially faster than classical computers~\cite{shor_algorithms_1994}, due to quantum processing. Grover's algorithm, on the other hand, reduces the complexity of unstructured search problems, enabling more efficient brute-force attacks against cryptographic algorithms~\cite{grover_fast_1996}. Specifically, Grover's algorithm reduced the time complexity of the brute-force attack from $O(N)$ to $O(\sqrt{N})$~\cite{durmuth_towards_2021}. 

These quantum threats have driven the development of post-quantum cryptography (PQC), particularly with the recent publication of the NIST PQC standards\cite{computer_security_division_post-quantum_2017, alagic_status_2025}. This development has also prompted discussions among quantum cryptography researchers about using Grover's algorithm to assess the quantum resistance of current cryptographic schemes. Consequently, considerable focus has been placed on designing and implementing quantum circuits for algorithms like AES, along with accurate resource estimation ~\cite{grassl_applying_2016}. Such implementations provide crucial insights into the quantum cost of cryptographic algorithms and enable accurate assessments of Grover's search attacks against these systems. In this context, quantum-cost refers to the total number of qubits and quantum gates required for implementation. However, the resource requirements of full-scale quantum attacks far exceed the capabilities of current noisy intermediate-scale quantum (NISQ) devices~\cite{jaques_implementing_2020}.

Beyond these limitations, lightweight cryptographic primitives such as Mini-AES provide valuable testbeds for studying the interaction between quantum circuit design and cryptanalysis under realistic resource constraints. Their simplified structure clarifies how algebraic operations map to quantum circuits while highlighting key optimization strategies, including ancilla reuse, mid-circuit measurement, and efficient gate decomposition. As a result, compact implementations serve both as pedagogical tools and as practical bridges between theoretical quantum cryptanalysis and future applications in resource-constrained environments such as the Internet of Things (IoT).

\subsection{Related Work} 
Block ciphers can be efficiently implemented as quantum circuits using bit-level reversible logic~\cite{luo_quantum_2024, jang_efficient_2021, jang_grover_2021}, making them natural targets for circuit-level quantum cryptanalysis~\cite{yadav_practical-quantum_2025}. In particular, lightweight ciphers have attracted much attention, with several designs already mapped to reversible circuits. Jang et al.~\cite{jang_efficient_2021} presented optimized quantum implementations of the PRESENT and GIFT block ciphers, requiring 144 and 192 logical qubits, respectively, and achieving T-depths of 311 and 308 using the Lighter-R S-box construction tool. Their Grover-resource estimates report approximately 2108 Toffoli, 4683 CNOT, and 1118 X gates for PRESENT, and 1792 Toffoli, 1792 CNOT, and 3261 X gates for GIFT. Luo et al.~\cite{luo_quantum_2024} investigated the SM4 cipher using composite-field reversible S-boxes and achieved a full-cipher implementation with only 260 qubits through circuit serialization. Jang et al.~\cite{jang_grover_2021} also demonstrated a compact two-round Simplified AES design requiring 32 qubits, with T-depth 47, 144 CNOTs, 96 Toffoli, and 16 triple-controlled Toffoli gates. Beyond lightweight designs, substantial effort has focused on quantum implementations of the AES S-box. Grassl et al.~\cite{grassl_applying_2016} observed that SubBytes is a permutation and constructed a reversible circuit requiring only 9 qubits but with high gate cost (1385 Toffoli and 1551 CNOT/NOT gates). Exploiting the algebraic structure of SubBytes, they proposed an alternative construction mapping $\lvert x_i \rangle \lvert 0 \rangle_{32}$ to $\lvert x_i \rangle \lvert S(x_i) \rangle \lvert 0 \rangle$, using 40 qubits and reducing the cost to 512 Toffoli, 469 CNOT, and 4 NOT gates. Building on this work, Kim et al.~\cite{kim_timespace_2018} studied time–space trade-offs and parallelization strategies for Grover-based AES key search, while Langenberg et al.~\cite{langenberg_reducing_2020} reduced the AES-128 Toffoli count by more than 88\%. Further optimizations eliminated one $\mathbb{F}_{2^8}$ multiplication, and Almazrooie et al.~\cite{almazrooie_quantum_2018} reduced the Toffoli count to 448 using 56 qubits, together with 494 CNOT and 4 NOT gates~\cite{boyar_new_2010}. 

In this paper, we present a quantum implementation of Mini-AES, a lightweight cipher well suited for exploring AES fundamentals in resource-constrained research environments. Unlike standard AES, which operates over GF($2^8$), Mini-AES works in GF($2^4$), simplifying the underlying arithmetic while preserving the core round structure. Field operations in GF($2^4$) can be expressed in algebraic normal form (ANF), where addition and multiplication correspond to XOR and AND operations, respectively. In quantum circuits, these map directly to CNOT and three-qubit Toffoli gates, enabling efficient bit-level implementations of key scheduling, permutation, and substitution operations. This minimal yet non-trivial structure makes Mini-AES an effective testbed for quantum-circuit optimization and Grover-based cryptanalysis~\cite{mondal_improved_2020}. Since Mini-AES is a 16-bit pedagogical cipher, its most direct point of comparison is other 16-bit AES-like designs such as Simplified/Sample-AES~\cite{jang_grover_2021, wang_quantum_2022}, as summarized in Table~\ref{com}.

% \begin{table*}[t]
% \caption{Summary of quantum gates: Pauli-X, CNOT, and Toffoli gates with matrix representations.}
% \label{gates_sum}
% \centering
% \scriptsize
% % \renewcommand{\arraystretch}{0.7} % Reduce row height
% % \setlength{\tabcolsep}{4pt} % Reduce column separation
% \begin{tabular}{|c|c|c|p{3.5cm}|p{2.5cm}|}
% \hline \rowcolor{lightgray}
% \textbf{Gate} & \textbf{Symbol} & \textbf{Matrix} & \textbf{Description} & \textbf{Image} \\
% \hline
% Pauli-X & $X$ & $\begin{bmatrix}0 & 1 \\ 1 & 0\end{bmatrix}$ & 
% Flips qubit state (NOT gate) & 
% \raisebox{-0.45\height}{\includegraphics[width=2cm]{Graphics/Xgate.png}} \\
% \hline
% CNOT & $CX$ & $ \begin{bmatrix}1 & 0 & 0 & 0 \\ 0 & 1 & 0 & 0 \\ 0 & 0 & 0 & 1 \\ 0 & 0 & 1 & 0\end{bmatrix}$ & 
% Flips target if control is $\lvert 1 \rangle$ & 
% \raisebox{-0.45\height}{\includegraphics[width=2cm]{Graphics/CX.png}} \\
% \hline
% Toffoli & $CCX$ & $\begin{bmatrix}1 & 0 & 0 & 0 & 0 & 0 & 0 & 0 \\ 0 & 1 & 0 & 0 & 0 & 0 & 0 & 0 \\ 0 & 0 & 1 & 0 & 0 & 0 & 0 & 0 \\ 0 & 0 & 0 & 1 & 0 & 0 & 0 & 0 \\ 0 & 0 & 0 & 0 & 1 & 0 & 0 & 0 \\ 0 & 0 & 0 & 0 & 0 & 1 & 0 & 0 \\ 0 & 0 & 0 & 0 & 0 & 0 & 0 & 1 \\ 0 & 0 & 0 & 0 & 0 & 0 & 1 & 0\end{bmatrix}$ & 
% Flips target if both controls are $\lvert 1 \rangle$ &
% \raisebox{-0.45\height}{\includegraphics[width=2cm]{Graphics/CCX.png}} \\ 
% \hline
% \end{tabular}
% \end{table*}

\subsection{Contribution}
This work presents a compact and resource-efficient quantum circuit design for Mini-AES encryption, specifically optimized for educational platforms and classical quantum simulators. We implement the complete Mini-AES encryption process using quantum logic in Qiskit, translating S-box operations through algebraic normal form (ANF) equations and mapping them to CNOT and Toffoli-3 gates. Our main contributions are as follows:
\begin{itemize}
    \item A qubit-reuse strategy based on reset operations, enabling full Mini-AES encryption execution with only 28 qubits.
    \item A resource-optimized S-box implementation using 6 qubits only: 4 qubits for input state representation (nibbles) and 2 ancilla qubits, which are systematically reset after each ANF output transfer.
    \item A design where ancilla qubits are reset and outputs stored in dedicated quantum registers, supporting subsequent Shift-Rows and Mix-Columns operations.
    \item An open-source Python tool that automatically generates Qiskit circuit blocks from ANF equations, facilitating adaptation to other lightweight cryptographic algorithms~\cite{shahmir_resource-constrained_2025}.
    \item Detailed simulation-based evaluation with gate counts, T-gate depth, and system performance metrics, providing precise figures for Grover-oracle cost estimation, including oracle depth, qubit–time complexity, and scalability analyses.
\end{itemize}

\subsection{Organization}
The rest of the paper is organized as follows. Section~\ref{Preliminaries} gives an understanding of the work by introducing preliminaries. Section~\ref{Mini-AES} reviews the Mini-AES cipher along with its core operations. Section~\ref{Circuit_Design} and Section~\ref{Circuit_Implementation} present our ANF-based circuit design methodology and its implementation in Qiskit. Section~\ref{Results} reports the experimental validation and resource analysis, including metrics relevant to Grover-oracle cost estimation. Finally, Section~\ref{Conclusion} summarizes the contributions and outlines directions for future work.

\section{Preliminaries}\label{Preliminaries}
Grover’s algorithm is a quantum search procedure designed to locate a specific element within an unsorted or unstructured database faster than classical methods. While a classical search requires $O(N)$ queries for a database of size $N$, Grover’s algorithm achieves the task in approximately $O(\sqrt{N})$ steps, offering a provably optimal quadratic speedup. Although not exponential, this improvement demonstrates how quantum mechanics can outperform classical computation in specific problem domains. Originally proposed by Lov Grover in 1996~\cite{grover_fast_1996}, the algorithm has become a cornerstone of quantum computing. However, its speedup is limited: for NP-complete problems, which typically require exponential time classically, Grover’s algorithm still demands exponential resources, albeit reduced by a quadratic factor.Although practical implementations of Grover’s algorithm require large-scale fault-tolerant quantum computers and remain beyond current technological capabilities, its theoretical implications have already influenced cryptographic research. In cryptanalytic applications, Grover’s algorithm can be applied to symmetric key cryptosystems by performing an exhaustive key search. For a cipher with a $k$-bit key, classical brute-force attacks require an average of $2^{k-1}$ operations, whereas Grover’s algorithm reduces this to approximately $2^{k/2}$ quantum operations. This effectively halves the security level of symmetric cryptographic primitives: AES-128 provides 64 bits of quantum security, AES-192 provides 96 bits, and AES-256 provides 128 bits of post-quantum security. Understanding Grover’s algorithm is therefore essential for evaluating the quantum resistance of cryptographic protocols and guiding the design of systems secure in the post-quantum era. While current hardware limitations provide a temporary grace period, the algorithm’s theoretical impact continues to drive the development of quantum-resistant cryptographic standards and implementation strategies.

\section{Mini-AES}\label{Mini-AES}
In 2000, NIST selected the Rijndael algorithm as the Advanced Encryption Standard (AES), later standardized as FIPS~197, establishing it as the dominant symmetric-key cipher worldwide. AES operates on 128-bit blocks with key sizes of 128, 192, or 256 bits, providing corresponding classical security levels.
Under quantum attack models based on Grover’s algorithm, this security is reduced by a quadratic factor~\cite{bonnetain_quantum_2019}. AES derives its strength from a substitution–permutation network applied over multiple rounds, combining SubBytes, ShiftRows, MixColumns, and AddRoundKey operations defined over GF($2^8$). While this design offers strong security margins, the underlying GF($2^8$) arithmetic and complex key schedule introduce substantial overhead for educational use and quantum circuit simulation.
\begin{figure}[b]
    \centering
    \includegraphics[scale=0.27]{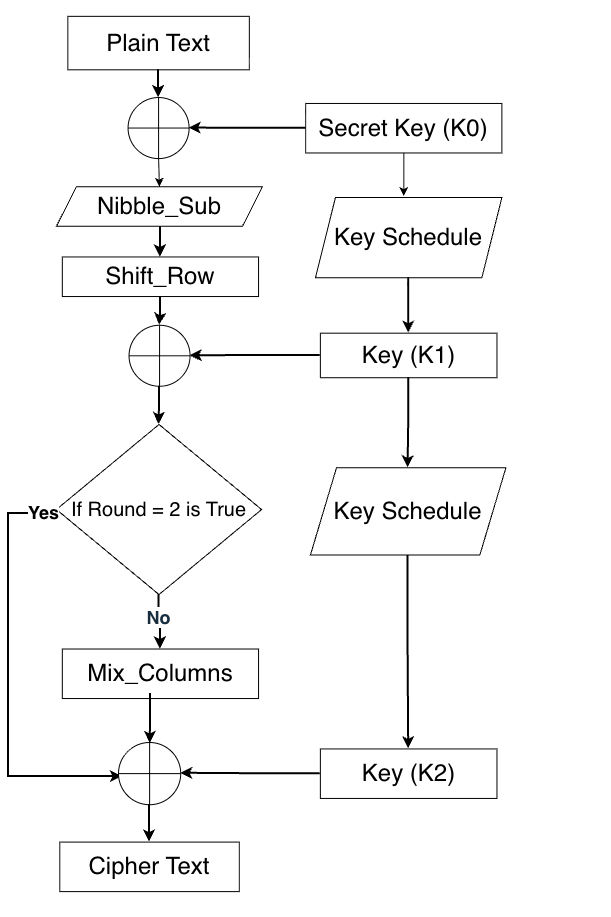}
    \caption{Flow chart representation of Mini-AES.}
    \label{mini-AES}
\end{figure}
To address these limitations while preserving the AES round structure, Mini-AES was introduced as a pedagogically oriented 16-bit block cipher~\cite{phan_mini_2002}. Mini-AES mirrors the core AES operations but operates over GF($2^4$) using the irreducible polynomial $x^4+x+1$, employs a simplified key schedule, and executes only two rounds. The complete two-round Mini-AES encryption process can therefore be expressed compactly as:
\begin{equation}
     C = (K_2 \oplus \pi \oplus \gamma \oplus K_1 \oplus \theta \oplus \pi \oplus \gamma \oplus K_0) \oplus P,
\end{equation}
where: $C$ is the ciphered text, $P$ is the plain text, $K_0$, $K_1$, and $K_2$ are key schedules, $\gamma$ is Nibble-Sub, $\theta$ is Mix-Column, and $\pi$ is Shift-Row. Similarly, for decryption, we have:
\begin{equation}
    P = (K_{0} \oplus  \gamma^{-1} \oplus \pi \oplus \theta \oplus K_{1} \oplus  \gamma^{-1} \oplus \pi \oplus K_{2}) \oplus C.
\end{equation}
Here, $\gamma^{-1}$  is the inverse Nibble-Sub, while other operations are similar but in reverse order. In the following, we describe the basic operations: Nibble-Sub, Mix-Column, Shift-Row, and Key XOR.
\paragraph{Nibble Sub ($\gamma$)} The Nibble substitution operation provides the primary source of nonlinearity in Mini-AES by applying a fixed 4-bit to 4-bit substitution to each nibble of the 16-bit state. The complete state is partitioned into four nibbles, with each 4-bit segment independently transformed through a predetermined substitution box (S-box) as defined in Fig.~\ref{fig:tables}. This nonlinear transformation serves as the confusion component of the cipher, obscuring the mathematical relationship between the plaintext and ciphertext. Due to its critical role in the cipher's security and the complexity of its quantum implementation, the detailed quantum circuit design for the Nibble-Sub operation is presented comprehensively in Section~\ref{Circuit_Design}.
\begin{figure}
\centering
\scriptsize
      \begin{subfigure}[b]{0.2\textwidth}
        \centering 
        \resizebox{0.5\textwidth}{!}{
        \begin{tabular}{|c|c|}\hline \rowcolor{lightgray}
            \textbf{Input} & \textbf{Output} \\\hline
            0000 & 1110  \\\hline
            0001 & 0100  \\\hline
            0010 & 1101  \\\hline 
            0011 & 0001  \\\hline
            0100 & 0010  \\\hline
            0101 & 1111  \\\hline
            0110 & 1011  \\\hline
            0111 & 1000  \\\hline 
            1000 & 0011  \\\hline
            1001 & 1010  \\\hline
            1010 & 0110 \\\hline
            1011 & 1100 \\\hline
            1100 & 0101 \\\hline
            1101 & 1001 \\\hline
            1110 & 0000 \\\hline
            1111 & 0111 \\\hline
        \end{tabular}}
        \caption{NibbleSub}
        \label{tab:table_a}
    \end{subfigure}
    \begin{subfigure}[b]{0.2\textwidth}
        \centering  
        \resizebox{0.5\textwidth}{!}{
        \begin{tabular}{|c|c|}\hline \rowcolor{lightgray}
            \textbf{Input} & \textbf{Output}  \\\hline
            0000 & 1110  \\\hline
            0001 & 0011   \\\hline
            0010 & 0100   \\\hline
            0011 & 1000  \\\hline
            0100 & 0001   \\\hline
            0101 & 1100   \\\hline
            0110 & 1010  \\\hline
            0111 & 1111   \\ \hline
            1000 & 0111 \\ \hline
            1001 & 1101 \\ \hline 
            1010 & 1001 \\ \hline
            1011 & 0110 \\ \hline
            1100 & 1011 \\ \hline
            1101 & 0010 \\ \hline
            1110 & 0000 \\ \hline
            1111 & 0101 \\ \hline
        \end{tabular}}
        \caption{Inverse NibbleSub}
        \label{tab:table_b}
    \end{subfigure}
     \caption{Nonlinear substitutions in Mini-AES implementations~\cite{phan_mini_2002}.}
    \label{fig:tables}
\end{figure}
\paragraph{Mix-Column ($\theta$)}
The mix-column operation performs a linear transformation that provides diffusion by mixing the nibbles within each column of the state matrix. In Mini-AES, this operation is implemented as matrix multiplication over GF($2^4$), where the state vector is multiplied by a fixed $2 \times 2$ circulant matrix. This transformation ensures that changes in input bits propagate across multiple output positions, contributing to the cipher's avalanche effect. The mix-column operation for Mini-AES can also be expressed in the form of equations~\cite{phan_mini_2002}, which is written as,
\begin{align}
    d_0 = 3 c_0 \oplus 2 c_1 \label{eq:d_0}\\
    d_1 = 2 c_0 \oplus 3 c_1 \label{eq:d_1}\\
    d_2 = 3 c_2 \oplus 2 c_3 \label{eq:d_2}\\
    d_3 = 2 c_2 \oplus 3 c_3\label{eq:d_3}.
\end{align}
Here, $[c_0, c_1, c_2, c_3]$ represents the input state of bits and $[d_0, d_1, d_2, d_3]$ represents the output bits after permutation. Also, we can see that the multiplication of 3, i.e. $0011$, and 2, i.e. $0010$, needs to be implemented on specified bits to apply the mix-column completely. The quantum version of this multiplier is given in Fig.~\ref{fig:combined}a and Fig.~\ref{fig:combined}b. While the complete implementation of the mix-column is given in Fig.~\ref{fig:combined}c, which uses the multiplier circuits and their inverse as well.

\paragraph{Shift-Row ($\pi$) and Key-Addition}
The key qubit states are placed directly in front of the plaintext qubits, allowing an immediate XOR between the live qubit states and the key without requiring a dedicated key-schedule register. This approach is more resource-efficient for short-term operations. Alternatively, a separate 16-qubit register can be prepared for the round key, with CNOT gates used to XOR it with the plaintext register. In our implementation, we adopt the first approach to minimize qubit usage and simplify the circuit design. 

\begin{figure*}[t]
    \centering
    \begin{subfigure}[b]{1\textwidth}
        \centering
        \includegraphics[scale=0.3]{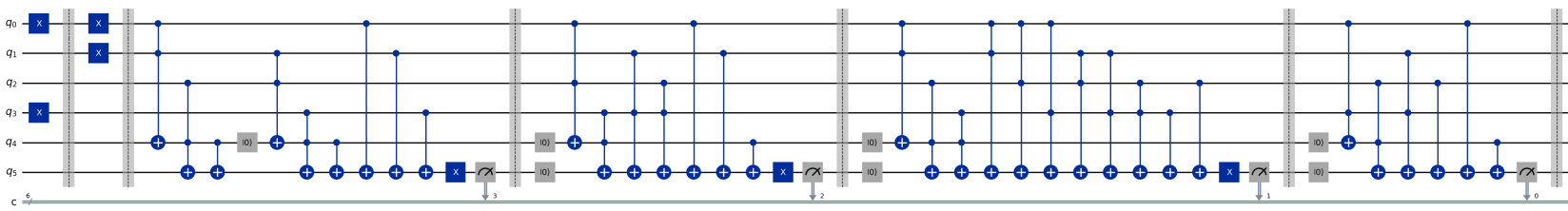}
        \caption{Nibble-Sub used for encryption}
        \label{fig:Sbox1a}
    \end{subfigure}
    \hfill\\
    \begin{subfigure}[b]{1\textwidth}
        \centering
        \includegraphics[scale=0.202]{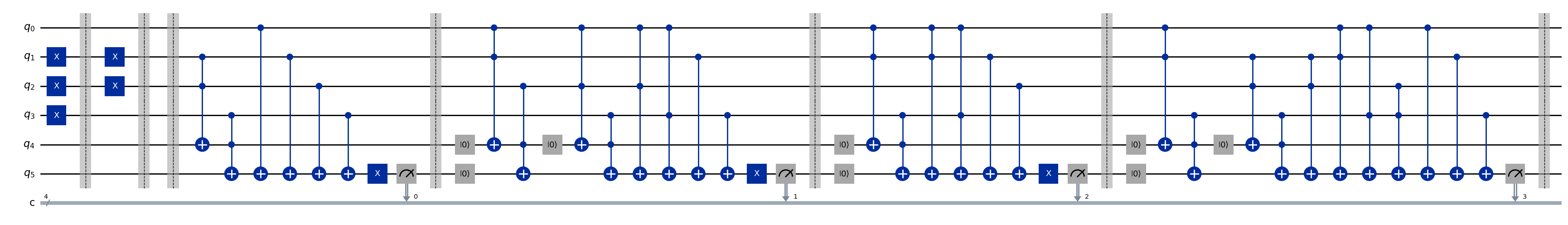}
        \caption{Inverse Nibble-Sub used for decryption}
        \label{fig:Sbox1b}
    \end{subfigure}
    \caption{Schematics of the Quantum circuit of the Mini-AES S-box (Nibble-Sub) using Toffoli and CNOT gates derived from ANF expressions.}
    \label{fig:Sbox1}
\end{figure*}

\section{Circuit Design}\label{Circuit_Design}
To enable algebraic analysis of Mini-AES's nonlinear components, we employed computational tools to derive the Algebraic Normal Form (ANF) representation of the Nibble-Sub operation. While comprehensive systems like SageMath provide extensive finite field arithmetic capabilities, we propose and develope a specialized, standalone implementation~\cite{shahmir_resource-constrained_2025} that efficiently converts the 4-bit S-box lookup table into polynomial equations over GF(2). This implementation systematically processes all 16 input-output pairs to generate the minimal polynomial representation for each output bit, thereby expressing the S-box's nonlinear transformation as a system of Boolean polynomials essential for subsequent circuit design and cryptanalytic applications. The resulting ANF equations for the Mini-AES Nibble-Sub operation are given as:
\begin{align}
    f(x) = \sum_{y \epsilon \mathcal{F}_{2^n}} f_y x^y,
\end{align}
where $f_y$ are the inverse coefficients of the Nibble-Sub output entries and $x^y$ is an algebraic integer that can be represented in terms of binary input bits. By substituting the values of polynomials, i.e., the output of the Nibble-Sub \{14,\allowbreak 10,\allowbreak 3,\allowbreak 6,\allowbreak 12,\allowbreak 
7,\allowbreak 10,\allowbreak 8,\allowbreak 13,\allowbreak 3,\allowbreak 6,\allowbreak 5,\allowbreak 10,\allowbreak 2,\allowbreak 10,\allowbreak 0\}, also given in Fig.~\ref{tab:table_a} in binary form, the ANF equations for the given Nibble-Sub vector are written as:
\begin{align}
y_0 &= x_1  \oplus  x_0  x_2 \oplus x_3 \oplus x_0  x_3 \oplus x_0  x_1  x_3  \label{eq:y_0}\\ 
\begin{split}
 y_1 &= 1 \oplus x_0 \oplus x_1 \oplus x_0   x_1 \oplus x_0   x_2 \oplus x_1   x_2 \oplus x_0   x_3 \\ &\qquad \oplus x_1   x_3 \oplus x_2   x_3 \oplus x_0   x_2   x_3 \oplus x_1   x_2   x_3        
\end{split}  \label{eq:y_1}\\
y_2 &= 1 \oplus x_0   x_1 \oplus x_2 \oplus x_0   x_2 \oplus x_3 \oplus x_1   x_3 \oplus x_0   x_1   x_3  \label{eq:y_2}\\
y_3 &= 1 \oplus x_0 \oplus x_2 \oplus x_1   x_2 \oplus x_0    x_1   x_2 \oplus x_3 \oplus x_2   x_3 \oplus x_1   x_2   x_3 ,  \label{eq:y_3}
\end{align}
where, $\{ x_0,x_1,x_2,x_3 \}$ are the input bits for the Nibble-Sub operation, and $\{ y_0,y_1,y_2,y_3 \}$ are the output bits. Here $\oplus$ represents the XOR operation, and the adjacent bits are operated with AND together. These operations are converted to a quantum circuit by using CNOT-gate for XOR operation and Toffoli-3 for AND operation. The final output of each ANF equation can be stored in a single target qubit. In our case, the target qubit is the sixth qubit ($q_5$ in Fig.~\ref{fig:Sbox1}).
Note that the target qubit must be in state $\ket{0}$, i.e., the low-energy state of the qubit. In addition, for experimental setups, this gives a much-reduced probability of bit-flips or decoherence to higher vibrational modes and makes sure that no extra values are added to the result of ANF when in the quantum circuit. Since the use of ANF description is fixed at design time, it keeps the constructed quantum circuit fixed for all problem instances. Whereas the overall encryption process remains instance-dependent due to varying plaintexts and round key additions. Finally, for the inverse Nibble-Sub operation, the inverse ANF expressions are derived analogously, using the corresponding output vector from Fig.~\ref{tab:table_b}. In the following section, we see the quantum implementation of Nibble-Sub using these ANF equations.

\section{Circuit Implementation}\label{Circuit_Implementation}
\begin{figure*}
\centering
\begin{minipage}[t]{0.48\textwidth}
    \centering
    \vspace{0pt} 
    \includegraphics[width=0.6\linewidth]{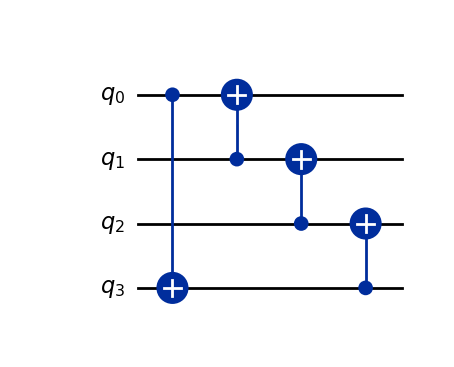} \label{3-Multiplier} \\
    (a) 3-Multiplier
    
    \includegraphics[width=0.6\linewidth]{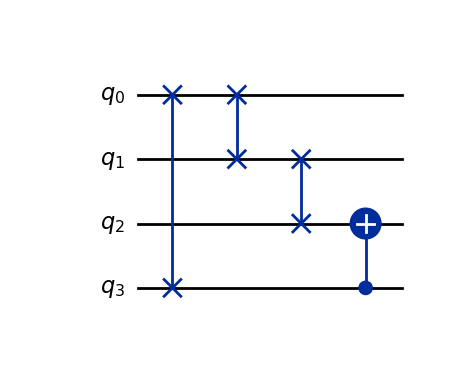} \label{2-Multiplier}\\
    (b) 2-Multiplier
\end{minipage}
\begin{minipage}[t]{0.48\textwidth}
    \centering
    \vspace{0pt} % align top
    \includegraphics[width=0.85\linewidth]{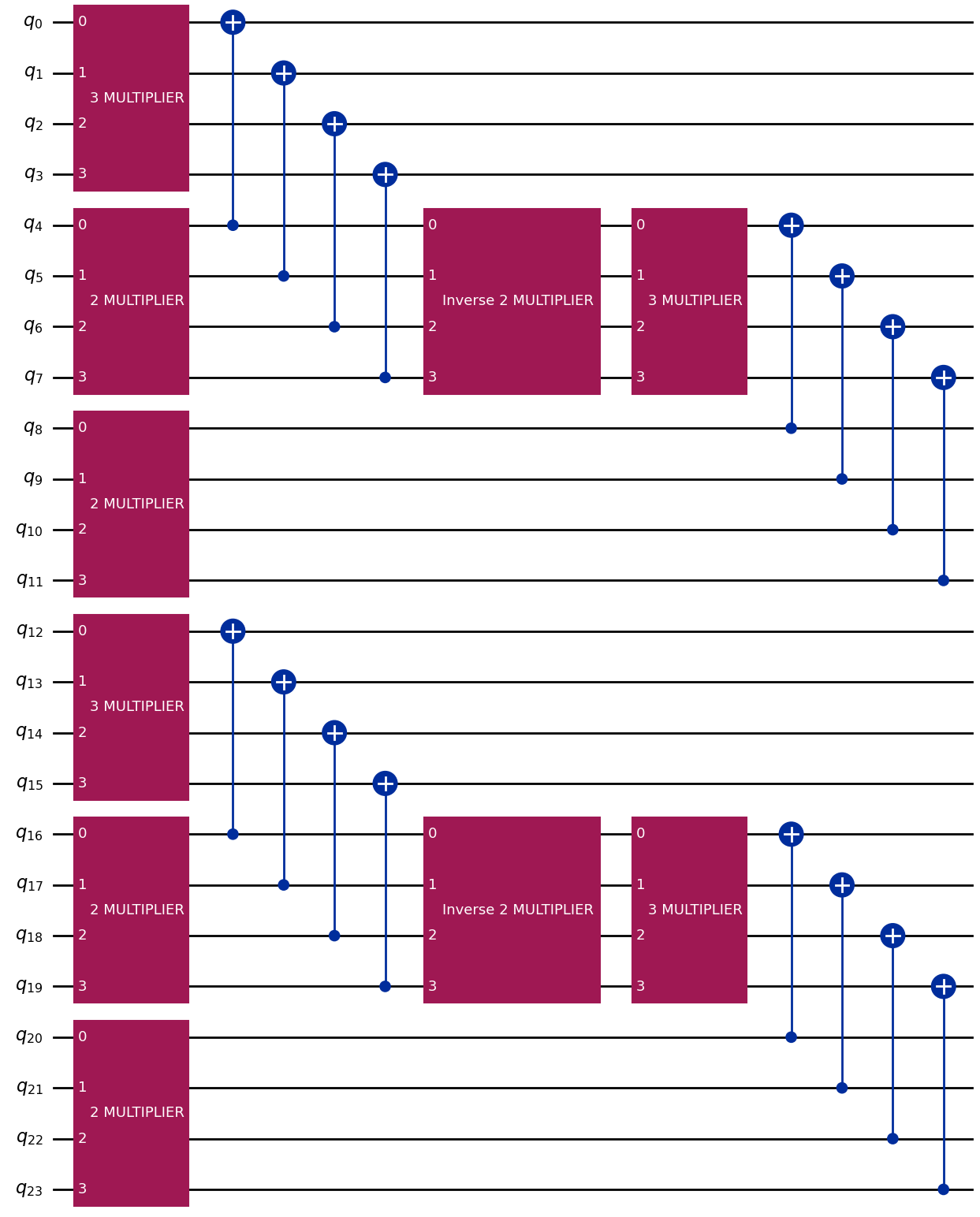} \\
    (c) Mix-Column operation using above multipliers
     % fill extra space to match height
\end{minipage}
\caption{Quantum circuit diagrams for mix-column operation, where the multiplier gates i.e. (a) 3 multiplier, (b) 2 multiplier are constructed using composite logic, and used in (c) the mix-column operation utilizing these multipliers and their inverse. Which are the same operations applied in reverse order.}
\label{fig:combined}
\end{figure*}

The core of the Mini-AES quantum implementation lies in efficiently constructing the nonlinear substitution layer, i.e., Nibble-Sub, using algebraic normal form (ANF) equations. Each ANF equation expresses a single output bit $y_i$ as a Boolean polynomial over input bits $\{x_0, x_1, x_2, x_3\}$ using XOR and AND operations. These are directly translated into quantum operations using CNOT (for XOR) and Toffoli-3 (for AND) gates. To aid reproducibility and clarify how the logic is mapped into the circuit, we present the pseudocode in Algorithm~\ref{algorithm1}.
\begin{algorithm}
\caption{ANF Term to Qiskit Circuit}\label{algorithm1}\scriptsize
\begin{algorithmic}[1]
\Function{ANFToQiskit}{$term$, $target$}
    \If{$term$ is a single variable $x_i$}
        \State \texttt{qc.cx($i$, $target$)}
    \ElsIf{$term$ is a product of two variables $x_i x_j$}
        \State \texttt{qc.ccx($i$, $j$, $target$)}
    \Else
        \State Partition $term$ into $(x_i x_j)$ and $x_k$
        \State \texttt{qc.ccx($i$, $j$, ancilla[0])}
        \State \texttt{qc.ccx($k$, ancilla[0], ancilla[1])}
        \State Reset ancilla[0] to $\ket{0}$
        \State Copy ancilla[1] to Memory Qubit
        \State Reset ancilla[1] 
    \EndIf
\EndFunction
\end{algorithmic}
\end{algorithm}
The Mini-AES consists of two rounds, where the implementation is divided into two parts: (i) Nibble-Sub and (ii) Permutations (shift-row and mix-column). The implementation of Nibble-Sub in each round takes 6 qubits: four to hold the state of the input register, and two ancilla qubits for evaluating the ANF equations. After each equation is computed, the ancilla qubits are reset, which reduces the overall quibit requirement since the computation proceeds sequentially. The schematics of the Nibble-Sub and its inverse are given in Fig.~\ref{fig:Sbox1}, where the operations before each measurement correspond to the ANF equations in \eqref{eq:y_0}-\eqref{eq:y_3}. 
Because subsequent operations require all Nibble-Sub outputs to remain physically active, the results are stored in an additional 16-qubit register that serves as a quantum memory. This uses qubits [6:21], while the original six qubits [0:5] used during NibbleSub are reset to save resources. In total, this implementation requires only 24 qubits for the first round and 187 gates after rearranging the qubit states. 
Following the Nibble-Sub, the Shift-Row operation is applied using four swap gates between the second and last nibbles, applied on the 16 memory qubits. The Mix-Column operation is then implemented through multiplier gates for multiplication by 2 and 3 (integers), designed using combinational logic. The construction of these gates is illustrated in Fig.~\ref{fig:combined}a and Fig.~\ref{fig:combined}b, respectively. 
These multipliers act on the 16-qubit Nibble-Sub output (after Shift-Row), mapped onto the qubit [8:16]. The first 8 qubits [0:7], which have been reset, are reused to store two output nibbles produced by the multiplier gates, as shown in Fig.~\ref{fig:combined}c. Using CNOT (XOR) operations with corresponding multiplier outputs, the states defined in \eqref{eq:d_0}-\eqref{eq:d_3} are fully mapped on the qubits [0:7] and [12:19], as depicted in Fig.~\ref{fig:combined}c. At this point, the first round of the Mini-AES is completed. 
\begin{algorithm}
\caption{Quantum Encryption Circuit}\scriptsize
\label{algo:enc}
\begin{algorithmic}[1]
\State Initialize Quantum Circuit with 28 qubits, 16 classical bits
\State Let $P = [P_0, P_1, P_2, P_3]$, $K = [K_0, K_1, K_2]$
\State Let $K_0$, $K_1$ and $K_2$ be round keys;
\State Let control variables be $s=8$; 

\For{$i = 0$ to $3$}
    \State Apply Basis Encoding for $P[i]$ followed by $K_0$ on \Call{qubits}{[0:15]}.
    \For{$j = 0$ to $3$}
        \State \Call{Algorithm 1}{}, \Call{qubits}{[0:5], $s$}, $s \gets s + 1$
        \State \Call{copy qubits}{[0:7]} to \Call{Memory qubits}{\empty}
    \EndFor
    \State \Call{reset qubits}{[0:7]}
\EndFor

\State \Call{Row\_Shift}{qc.swap([12:15],[20:23])}, 
\State Apply \Call{Mix\_Column}{QUBITS [0:23]}, \Call{reset qubits}{[8:11]}
\State Rearrangement: \Call{}{qc.cx([16:19],[8:11])}, 
\State Basis Encode $K_1$ on \Call{qubits}{[0:15]}, 
\State \Call{reset qubits}{[16:27]},
\State  Rearrangement: \Call{}{qc.cx([4:15],[16:27])}, \Call{reset}{4:15}

\For{$i = 0$ to $3$}
    \For{$j = 0$ to $3$}
        \State \Call{Algorithm 1}{}, \Call{qubits}{[0:5], $s$}, $s \gets s + 1$
    \EndFor
    \State \Call{reset qubits}{[0:7]}
    \State Let $t=16$;
    \If{$i < 3$}
        \State Copy \Call{}{qc.cx([$t$,$t+1$,$t+2$,$t+3$], [0,1,2,3])}, 
        \State \Call{reset qubits}{[$t$,$t+1$,$t+2$,$t+3$]}, $t \gets t +4$
    \EndIf
\EndFor

\State \Call{Row\_Shift}{qc.swap([10:13],[18:21])}, \State Basis Encode $K_2$ on \Call{qubits}{[6:21]},
\State \Call{measure qubits}{[6:21]} $\rightarrow$ classical bits [0:15]
\end{algorithmic}
\end{algorithm}
For the second round, the qubits are rearranged into the first 16 slots to simplify subsequent operations. This is done by resetting only qubit [8:11] and using SWAP gates to relocate the third nibble into its position. Although not strictly necessary, this rearrangement facilitates a more structured implementation. After the third nibble is relocated, qubits [12:15] are reset, and the fourth nibble (originally on qubits [16:19]) is moved in that position using swap gates. Either SWAP or CNOT gates can be applied for this relocation, with equivalent results. However, in practice, CNOT gates achieve a threefold reduction in gate depth compared to SWAP gates~\cite{m_q_cruz_shallow_2024}. 
Once the rearrangement is complete, the round key $K_1$ is added by applying NOT gates (Pauli-X operations) across the register [0:15] using the basis encoding method. Following key addition, the register is reorganized: all quibits except the first nibble are shifted into the lower 12 positions, which are refreshed via the reset gates as described above. 
The first nibble then undergoes the Nibble-Sub operation. To manage this efficiently, a \textsc{For} loop is introduced so that qubit states are relocated into four newly allocated qubits. This brings the total qubit count to 28. For implementing the S-box of the second nibble, CNOT gates are used to transfer its state into the reset qubits [0:3], while qubits [5:11] are reset for reuse. 
\begin{table*}[t]
    \centering
    \scriptsize
    \caption{Test vectors for implementation of Mini-AES.}
    % \scalebox{0.87}{
    \begin{tabular}{|c|c|c|c|c|c|}\hline \rowcolor{lightgray}
         \textbf{Plaintext} & \textbf{Key} & \textbf{S-box\textsubscript{1}} & \textbf{Permutations\textsubscript{1}} & \textbf{S-box\textsubscript{2}} & \textbf{Ciphertext}   \\ \hline
         
         001 1100 0110 0011 & 100 0011 1111 0000 & 1111 0111 1010 0001 & 0000 1110 0011 1110 & 0001 0000 0101 0100 &  0111 0010 1100 0110   \\ \hline
         
          1111 0101 1010 1111 & 100 0011 1111 0000 & 0001 1011 1111 0111 & 1101 1011 0111 0011 & 0000 1100 0011 0101 &  0110 0011 1010 1010   \\ \hline

          001 1100 0110 0011 & 1111 0101 1101 1110 & 1110 1110 1000 0100 & 1001 0011 0100 0010 & 0000 0011 1101 0110 &  1111 0010 1111 1001   \\ \hline
    \end{tabular}
    % }
    \label{tab:testvector}
\end{table*}
\begin{table*}[t]
    \centering
    \scriptsize
    \caption{Circuit complexity of Mini-AES.}
    \begin{tabular}{|c|c|c|c|}\hline \rowcolor{lightgray}
\empty & \textbf{No. of Qubits} & \textbf{T-counts} & \textbf{T-depth}  \\ \hline
First S-box   & 6 & {\small CNOT = 52, Toffoli = 80, Not = 28, and Reset = 52} & 187 \\\hline
Round 1 & 24 & CNOT = 106, Toffoli = 80, NOT = 38, Reset = 72, and SWAP = 22 & 207 \\ \hline
Re-arrangement &   & CNOT = 12, and Reset= 12 & \\\hline
Second S-box  &  28  & CNOT = 198, Toffoli = 160, Reset = 156, NOT = 50, and SWAP = 22  &  395  \\ \hline
Round 1 \& 2 & 28 & CNOT = 198, Toffoli = 160, Reset = 156, NOT = 58, and SWAP = 26 & 397 \\ \hline
    \end{tabular}
    \label{tab:3}
\end{table*}
\begin{algorithm}
\caption{Quantum Decryption Circuit}\scriptsize
\label{algo:dec}
\begin{algorithmic}[1]
\State Initialize Quantum Circuit with 28 qubits, 16 classical bits
\State Let $C = [C_0, C_1, C_2, C_3]$, $K_2 = [K_2^{(0)}, K_2^{(1)}, K_2^{(2)}, K_2^{(3)}]$
\State Let $K_0$, $K_1$ and $K_2$ be round keys; $s \gets 8$, $p \gets 6$

\For{$i = 0$ to $3$}
    \State Basis Encode $C[i]$ and $K_2^{[i]}$;
    \For{$j = 0$ to $3$}
        \State Use Inverse ANF equations and \Call{Algorithm 1}{}, \Call{Qubits}{[0:5], $s$}, $s \gets s + 1$
    \EndFor
    \State \Call{reset qubits}{[0:7]}
\EndFor

\State \Call{Row\_shift}{qc.swap([12:15],[20:23])}, 
\State Basis Encode $K_1$ on \Call{qubits}{[8:24]}, 
\State Apply \Call{Mix\_Column}{QUBITS [0:23]}, \Call{reset qubits}{[8:11]}, 
\State Rearrangement: \Call{}{qc.cx([16:19],[8:11]}
\State \Call{Row\_shift}{qc.cx([4:7],[12:15])}, 
\State \Call{reset qubits}{[16:27]}
\State Copy: \Call{}{qc.cx([4:15],[16:27]}, \Call{reset qubits}{[4:15]}
\For{$i = 0$ to $3$}
        \State Use Inverse ANF equations and \Call{Algorithm 1}{}, \Call{qubits}{[0:5], $p$}, $p \gets p + 1$
    Let t=16;
    \If{$i < 3$}
        \State Copy \Call{}{qc.cx([$t$,$t+1$,$t+2$,$t+3$], [0,1,2,3])}, 
        \State \Call{reset qubits}{[$t$,$t+1$,$t+2$,$t+3$]}, $t \gets t +4$
    \EndIf
\EndFor

\State Basis Encode $K_0$ on \Call{qubits}{[6:21]}
\State \Call{measure qubits}{[6:21] $\rightarrow$ classical bits [0:15]}
\end{algorithmic}
\end{algorithm}
Similarly, after each S-box implementation, the next nibble is moved into the first four slots, which are reset along with a lower set of qubits to make space for new states to be stored. The output of the second Nibble-Sub is stored in the new reset qubits [6:21]. A Shift-Row operation is then applied by swapping the second nibble [10,11,12,13] with the fourth nibble [18,19,20,21] using SWAP gates. This step is followed by the third Key-Addition $K_2$, applied via basis encoding on qubits [6:21]. Finally, the ciphered output is measured. The complete encryption code is available on GitHub~\cite{shahmir_resource-constrained_2025}, while the pseudocode is given in Algorithm~\ref{algo:enc}. 
The decryption process in Mini-AES is the logical inverse of the encryption procedure, achieved by reversing the order of operations and applying the corresponding inverse transformations. Specifically, decryption begins with the Inverse Nibble-Sub (Fig.~\ref{fig:Sbox1b}), followed by the inverse permutation layers. As in encryption, decryption uses a 28-qubit register, but starts from the ciphertext $C=[C_0, C_1, C_2, C_3]$ and uses the round keys $K_2$, $K_1$ and $K_0$ in reverse order. The decryption routine proceeds as follows. Each nibble of the ciphertext and the round key $K_2$ is basis-encoded on qubits [0:3]. The inverse Nibble-Sub operation (Fig.~\ref{fig:Sbox1b}) is then applied on qubits [0:5] and stored on memory qubits [0:7] via CNOT gates. Then, ancilla qubits are reset to ensure clean reuse for the next nibble. Once all nibbles are processed, a rearrangement operation is performed by swapping qubits [12:15] with [20:23]. Next, the round key $K_1$ is encoded onto qubits [8:24], followed by a Mix-Column operation on qubits [0:23]. Next, qubits [8:11] are reset, and the states on [16:19] are mapped onto [8:11] using CNOT gates. A RowShift permutation is then applied on qubits [4:7] and [12:15], logically reversing the Shift-Row permutation from encryption process. At the end of this stage, round one concludes by resetting qubits [16:27] and copying the intermediate states from [4:15] into [16:27] to prepare for the next round. The same process is repeated in round two, as detailed in Algorithm~\ref{algo:dec}. Finally, the key $K_0$ is basis-encoded on qubits [6:21], completing the inverse key addition. At this stage, the fully decrypted plaintext is obtained by measuring qubits [6:21] into 16 classical bits. Detailed encryption and decryption quantum circuits are given in the code available on Github~\cite{shahmir_resource-constrained_2025}. Finally, to validate our design, we simulated the entire Mini‑AES workflow in Qiskit, confirming that each encryption–decryption cycle reproduces the input exactly. Resource utilization for these implementations is evaluated in the following section.
\section{Results}\label{Results}
The Mini-AES quantum circuit was implemented and simulated using Qiskit. Due to system resource constraints, a reduced version of the Mini-AES algorithm was utilized. To enhance computational efficiency, GPU acceleration was employed, thereby improving resource utilization. This optimization resulted in decreased RAM consumption, reduced processing time, and a significantly lower processor load. The performance metrics are summarized in Table~\ref{tab:1}.
\begin{table}[H]
\caption{System usage for different implementations.}
\label{tab:1}
    \centering
    \scriptsize
    \begin{tabularx}{\columnwidth}{|c|X|X|X|}\hline \rowcolor{lightgray}
         \textbf{System Specs} & \textbf{Plain Implementation} & \textbf{Constraint Implementation} & \textbf{Constraint Implementation (GPU)} \\\hline
         RAM usage & 40.9+ GB & 8.25 GB  & 2.987 GB \\\hline
         Time & N/A & 308.297s & 65.767s \\\hline
         Processor & N/A & 95.1\% & 14.2\% \\\hline
    \end{tabularx}
\end{table}
The implementations were validated by ensuring that encrypting and then decrypting the plaintext perfectly reproduced the original test vectors, with no loss of information, as shown in Table~\ref{tab:testvector}. Thereby confirming the logical reversibility of the construction and having established functional correctness, we now turn our attention to evaluating the computational resources and performance requirements of the proposed Mini-AES quantum circuits. A detailed resource analysis of the Mini-AES quantum circuit was conducted, focusing on gate counts and circuit depth for each major component. The analysis includes the number of qubits used, along with the counts of CNOT, Toffoli, NOT, Reset, and SWAP gates, as well as the overall T-depth, summarized in Table~\ref{tab:3}.

The gate counts in Table~\ref{tab:3} correspond to one complete Mini-AES encryption round (of the two rounds), including all Nibble-Sub, Mix-Columns, and key addition operations. When used inside Grover’s algorithm, the oracle must evaluate one encryption per iteration, followed by an inverse decryption to uncompute the ancilla states, effectively doubling the gate cost per iteration. For an $n$-bit key, Grover’s search requires approximately $\pi\cdot 2^{(n/2)}/4$ iterations. For Mini-AES (16-bit key), this corresponds to $\pi \cdot 2^{(16/2)}/4$ iterations. Using these results, we can derive theoretical estimates for Grover’s oracle based on the Mini-AES implementation represented here. The estimated total T-depth for a brute-force search over the key space is: $T_{\text{depth,Grover}} = (\pi/4) \cdot 2^{16/2} \times 397 $, where 397 is the total T-depth for one complete Mini-AES encryption/decryption. Similarly, the estimated total qubit usage  for the brute-force attack is: $(\pi\cdot 2^{16/2}/4)\cdot(28) $, presenting the cost of creating and utilizing the circuit multiplied by the number of required iterations. At the hardware-level, a SWAP gate is equivalent to three CNOT gates, while a Toffoli-3 gate is equivalent to six CNOT gates \cite{li_quantum_2022}. Using this decomposition, the total estimated T-count for Grover’s brute-force attack on Mini-AES can be expressed as~\cite{grassl_applying_2016}:
\begin{align}
T_{\text{count,Grover}} 
  &= \left( \frac{\pi \cdot 2^{\frac{16}{2}}}{4} \right) \times \big( 1236 + 58 \big) ,
\end{align}
where $58$ represents the number of NOT gates, and 1236 is the total CNOT-equivalent gate count for one full Mini-AES encryption. This value is obtained by decomposing 160 Toffoli gates into 960 CNOT gates ($160\times6$), 26 SWAP gates into 78 CNOT gates ($26\times3$), and adding the native 198 CNOT gates~\cite{m_q_cruz_shallow_2024}. This Mini-AES quantum circuit serves as a foundation for scaling to more complex block ciphers. For example, it can be extended to Simplified AES, where 28 qubits are required in the first round and a total of 32 qubits are used for complete encryption. Similarly, AES-128 can be implemented using the same methodology and code~\cite{shahmir_resource-constrained_2025}. However, for a single AES-128 round, the S-box implementation alone requires 54 qubits. Beyond this scale, the computational demands exceed the capability of available classical hardware, making full simulation infeasible on our current systems due to excessive RAM requirements.

\section{Conclusion}\label{Conclusion}
Estimating the cost of a Grover key search attack provides a measure of the strength of symmetric key algorithms, such as AES, against quantum computers. Such estimates require an efficient implementation of the encryption and decryption processes of the target algorithms. In this paper, we focused on the most simplified version of AES, known as Mini-AES, which we implemented efficiently as a quantum circuit, and used to derive theoretical estimates of Grover's key search cost. Our implementation achieves Mini-AES encryption using only 28 qubits, with a T-depth of 397 and optimized ancilla reuse, demonstrating feasibility on resource-constrained quantum platforms. This approach enables researchers with limited RAM and computational resources to implement AES in Qiskit and gain a basic understanding of Grover's attack on AES. Moreover, by employing the ANF approach for the S-box, we achieve an optimized and minimal implementation. Future work will explore additional optimization techniques to further reduce the cost of block cipher implementations and explore their applications in quantum network-based scenarios. 

\section*{Acknowledgment}

This publication was made possible by ARG first Cycle grant \# ARG01-0501-230053  from the Qatar National Research Fund (a member of Qatar Foundation). The findings herein reflect the work, and are solely the responsibility, of the authors

\bibliographystyle{ieeetr}
\bibliography{AES_paper}

@article{li_quantum_2022,
	title = {Quantum {Fredkin} and {Toffoli} gates on a versatile programmable silicon photonic chip},
	volume = {8},
	copyright = {2022 The Author(s)},
	issn = {2056-6387},
	url = {https://www.nature.com/articles/s41534-022-00627-y},
	doi = {10.1038/s41534-022-00627-y},
	language = {en},
	number = {1},
	urldate = {2026-02-11},
	journal = {npj Quantum Information},
	publisher = {Nature Publishing Group},
	author = {Li, Yuan and Wan, Lingxiao and Zhang, Hui and Zhu, Huihui and Shi, Yuzhi and Chin, Lip Ket and Zhou, Xiaoqi and Kwek, Leong Chuan and Liu, Ai Qun},
	month = sep,
	year = {2022},
	pages = {112},
}

@misc{shahmir_resource-constrained_2025,
	title = {Resource-{Constrained} {Quantum} {Implementation} and {Analysis} of {Mini}-{AES} {Cipher} {Code} ({Github})},
	copyright = {Apache-2.0},
	url = {https://github.com/SyedShahmirKazmi/Implement-Sbox-int-a-Qiskit-code-by-using-ANF-equations-},
	urldate = {2026-02-08},
	author = {Shahmir, Syed},
	month = mar,
	year = {2025},
	note = {original-date: 2024-06-12T10:01:40Z},
}

@inproceedings{boyar_new_2010,
	address = {Berlin, Heidelberg},
	title = {A {New} {Combinational} {Logic} {Minimization} {Technique} with {Applications} to {Cryptology}},
	isbn = {978-3-642-13193-6},
	doi = {10.1007/978-3-642-13193-6_16},
	language = {en},
	booktitle = {Experimental {Algorithms}},
	publisher = {Springer},
	author = {Boyar, Joan and Peralta, Ren{\textbackslash}'\{e\}},
	editor = {Festa, Paola},
	year = {2010},
	pages = {178--189},
}

@techreport{alagic_status_2025,
	title = {Status {Report} on the {Fourth} {Round} of the {NIST} {Post}-{Quantum} {Cryptography} {Standardization} {Process}},
	url = {https://csrc.nist.gov/pubs/ir/8545/final},
	doi = {10.6028/NIST.IR.8545},
	language = {en},
	number = {NIST Internal or Interagency Report (NISTIR) 8545},
	urldate = {2025-09-29},
	institution = {National Institute of Standards and Technology},
	author = {Alagic, Gorjan and Bros, Maxime and Ciadoux, Pierre and Cooper, David and Dang, Quynh and Dang, Thinh and Kelsey, John and Lichtinger, Jacob and Liu, Yi-Kai and Miller, Carl and Moody, Dustin and Peralta, Rene and Perlner, Ray and Robinson, Angela and Silberg, Hamilton and Smith-Tone, Daniel and Waller, Noah},
	month = mar,
	year = {2025},
}

@misc{computer_security_division_post-quantum_2017,
	title = {Post-{Quantum} {Cryptography} {\textbar} {CSRC} {\textbar} {CSRC}},
	url = {https://csrc.nist.gov/projects/post-quantum-cryptography},
	language = {EN-US},
	urldate = {2025-09-29},
	journal = {CSRC {\textbar} NIST},
	author = {Computer Security Division, Information Technology Laboratory},
	month = jan,
	year = {2017},
}

@article{kim_timespace_2018,
	title = {Time–space complexity of quantum search algorithms in symmetric cryptanalysis: applying to {AES} and {SHA}-2},
	volume = {17},
	issn = {1573-1332},
	shorttitle = {Time–space complexity of quantum search algorithms in symmetric cryptanalysis},
	url = {https://doi.org/10.1007/s11128-018-2107-3},
	doi = {10.1007/s11128-018-2107-3},
	language = {en},
	number = {12},
	urldate = {2025-09-03},
	journal = {Quantum Information Processing},
	author = {Kim, Panjin and Han, Daewan and Jeong, Kyung Chul},
	month = oct,
	year = {2018},
	pages = {339},
}

@article{wang_quantum_2022,
	title = {A quantum circuit design of {AES} requiring fewer quantum qubits and gate operations},
	volume = {17},
	issn = {2095-0470},
	url = {https://doi.org/10.1007/s11467-021-1141-2},
	doi = {10.1007/s11467-021-1141-2},
	language = {en},
	number = {4},
	urldate = {2025-09-03},
	journal = {Frontiers of Physics},
	author = {Wang, Ze-Guo and Wei, Shi-Jie and Long, Gui-Lu},
	month = feb,
	year = {2022},
	pages = {41501},
}

@article{sakhi_grover_2011,
	title = {Grover {Algorithm} {Applied} to {Four} {Qubits} {System}},
	volume = {4},
	copyright = {Copyright (c) 2011 Z. Sakhi,A. Tragha,R. Kabil,M. Bennai},
	issn = {1913-8989},
	url = {https://ccsenet.org/journal/index.php/cis/article/view/9055},
	doi = {10.5539/cis.v4n3p125},
	language = {en},
	number = {3},
	urldate = {2025-09-03},
	journal = {Computer and Information Science},
	author = {Sakhi, Z. and Tragha, A. and Kabil, R. and Bennai, M.},
	month = apr,
	year = {2011},
	pages = {p125},
}

@inproceedings{grover_fast_1996,
	address = {New York, NY, USA},
	series = {{STOC} '96},
	title = {A fast quantum mechanical algorithm for database search},
	isbn = {978-0-89791-785-8},
	url = {https://dl.acm.org/doi/10.1145/237814.237866},
	doi = {10.1145/237814.237866},
	urldate = {2025-07-02},
	booktitle = {Proceedings of the twenty-eighth annual {ACM} symposium on {Theory} of {Computing}},
	publisher = {Association for Computing Machinery},
	author = {Grover, Lov K.},
	month = jul,
	year = {1996},
	pages = {212--219},
}

@inproceedings{shor_algorithms_1994,
	title = {Algorithms for quantum computation: discrete logarithms and factoring},
	shorttitle = {Algorithms for quantum computation},
	url = {https://ieeexplore.ieee.org/document/365700},
	doi = {10.1109/SFCS.1994.365700},
	urldate = {2024-11-08},
	booktitle = {Proceedings 35th {Annual} {Symposium} on {Foundations} of {Computer} {Science}},
	author = {Shor, P.W.},
	month = nov,
	year = {1994},
	pages = {124--134},
}

@inproceedings{grassl_applying_2016,
	address = {Cham},
	title = {Applying {Grover}’s {Algorithm} to {AES}: {Quantum} {Resource} {Estimates}},
	isbn = {978-3-319-29360-8},
	shorttitle = {Applying {Grover}’s {Algorithm} to {AES}},
	doi = {10.1007/978-3-319-29360-8_3},
	language = {en},
	booktitle = {Post-{Quantum} {Cryptography}},
	publisher = {Springer International Publishing},
	author = {Grassl, Markus and Langenberg, Brandon and Roetteler, Martin and Steinwandt, Rainer},
	editor = {Takagi, Tsuyoshi},
	year = {2016},
	pages = {29--43},
}

@article{m_q_cruz_shallow_2024,
	title = {Shallow unitary decompositions of quantum {Fredkin} and {Toffoli} gates for connectivity-aware equivalent circuit averaging},
	volume = {1},
	issn = {2835-0103},
	url = {https://doi.org/10.1063/5.0187026},
	doi = {10.1063/5.0187026},
	number = {1},
	urldate = {2025-09-03},
	journal = {APL Quantum},
	author = {M. Q. Cruz, Pedro and Murta, Bruno},
	month = mar,
	year = {2024},
	pages = {016105},
}

@article{phan_mini_2002,
	title = {Mini {Advanced} {Encryption} {Standard} (mini-{Aes}): {A} {Testbed} for {Cryptanalysis} {Students}},
	volume = {26},
	issn = {0161-1194},
	shorttitle = {Mini {Advanced} {Encryption} {Standard} (mini-{Aes})},
	url = {https://doi.org/10.1080/0161-110291890948},
	doi = {10.1080/0161-110291890948},
	number = {4},
	urldate = {2025-09-03},
	journal = {Cryptologia},
	publisher = {Taylor \& Francis},
	author = {Phan, Raphael Chung-Wei},
	month = oct,
	year = {2002},
	note = {\_eprint: https://doi.org/10.1080/0161-110291890948},
	pages = {283--306},
}

@article{bonnetain_quantum_2019,
	title = {Quantum {Security} {Analysis} of {AES}},
	copyright = {Copyright (c) 2019 Xavier Bonnetain, María Naya-Plasencia, André Schrottenloher},
	issn = {2519-173X},
	url = {https://tosc.iacr.org/index.php/ToSC/article/view/8314},
	doi = {10.13154/tosc.v2019.i2.55-93},
	language = {en},
	urldate = {2025-09-03},
	journal = {IACR Transactions on Symmetric Cryptology},
	author = {Bonnetain, Xavier and Naya-Plasencia, María and Schrottenloher, André},
	month = jun,
	year = {2019},
	pages = {55--93},
}

@inproceedings{mondal_improved_2020,
	title = {An {Improved} {Online} {Testing} {Technique} {For} {Reversible} {Circuits}},
	url = {https://ieeexplore.ieee.org/document/9262971},
	doi = {10.1109/ISDCS49393.2020.9262971},
	urldate = {2025-09-03},
	booktitle = {2020 {International} {Symposium} on {Devices}, {Circuits} and {Systems} ({ISDCS})},
	author = {Mondal, Joyati and Das, Debesh Kumar},
	month = mar,
	year = {2020},
	pages = {1--5},
}

@article{yadav_practical-quantum_2025,
	title = {A practical-quantum differential attack on block ciphers},
	volume = {17},
	issn = {1936-2455},
	url = {https://doi.org/10.1007/s12095-023-00650-6},
	doi = {10.1007/s12095-023-00650-6},
	language = {en},
	number = {2},
	urldate = {2025-09-03},
	journal = {Cryptography and Communications},
	author = {Yadav, Tarun and Kumar, Manoj and Kumar, Amit and Pal, S. K.},
	month = mar,
	year = {2025},
	pages = {337--357},
}

@inproceedings{jang_grover_2021,
	title = {Grover on {Simplified} {AES}},
	url = {https://ieeexplore.ieee.org/document/9642017},
	doi = {10.1109/ICCE-Asia53811.2021.9642017},
	urldate = {2025-09-03},
	booktitle = {2021 {IEEE} {International} {Conference} on {Consumer} {Electronics}-{Asia} ({ICCE}-{Asia})},
	author = {Jang, Kyung-Bae and Song, Gyeong-Ju and Kim, Hyun-Ji and Seo, Hwa-Jeong},
	month = nov,
	year = {2021},
	pages = {1--4},
}

@article{jang_efficient_2021,
	title = {Efficient {Implementation} of {PRESENT} and {GIFT} on {Quantum} {Computers}},
	volume = {11},
	copyright = {http://creativecommons.org/licenses/by/3.0/},
	issn = {2076-3417},
	url = {https://www.mdpi.com/2076-3417/11/11/4776},
	doi = {10.3390/app11114776},
	language = {en},
	number = {11},
	urldate = {2025-09-03},
	journal = {Applied Sciences},
	publisher = {Multidisciplinary Digital Publishing Institute},
	author = {Jang, Kyungbae and Song, Gyeongju and Kim, Hyunjun and Kwon, Hyeokdong and Kim, Hyunji and Seo, Hwajeong},
	month = jan,
	year = {2021},
	pages = {4776},
}

@article{luo_quantum_2024,
	title = {Quantum circuit implementations of {SM4} block cipher optimizing the number of qubits},
	volume = {23},
	issn = {1573-1332},
	url = {https://doi.org/10.1007/s11128-024-04394-x},
	doi = {10.1007/s11128-024-04394-x},
	language = {en},
	number = {5},
	urldate = {2025-09-03},
	journal = {Quantum Information Processing},
	author = {Luo, Qing-bin and Li, Qiang and Li, Xiao-yu and Yang, Guo-wu and Shen, Jinan and Zheng, Minghui},
	month = may,
	year = {2024},
	pages = {177},
}

@inproceedings{jaques_implementing_2020,
	address = {Berlin, Heidelberg},
	title = {Implementing {Grover} {Oracles} for {Quantum} {Key} {Search} on {AES} and {LowMC}},
	isbn = {978-3-030-45723-5},
	url = {https://doi.org/10.1007/978-3-030-45724-2_10},
	doi = {10.1007/978-3-030-45724-2_10},
	urldate = {2025-09-03},
	booktitle = {Advances in {Cryptology} – {EUROCRYPT} 2020: 39th {Annual} {International} {Conference} on the {Theory} and {Applications} of {Cryptographic} {Techniques}, {Zagreb}, {Croatia}, {May} 10–14, 2020, {Proceedings}, {Part} {II}},
	publisher = {Springer-Verlag},
	author = {Jaques, Samuel and Naehrig, Michael and Roetteler, Martin and Virdia, Fernando},
	month = may,
	year = {2020},
	pages = {280--310},
}

@article{langenberg_reducing_2020,
	title = {Reducing the {Cost} of {Implementing} the {Advanced} {Encryption} {Standard} as a {Quantum} {Circuit}},
	volume = {1},
	issn = {2689-1808},
	url = {https://ieeexplore.ieee.org/document/8961201},
	doi = {10.1109/TQE.2020.2965697},
	urldate = {2025-09-03},
	journal = {IEEE Transactions on Quantum Engineering},
	author = {Langenberg, Brandon and Pham, Hai and Steinwandt, Rainer},
	year = {2020},
	pages = {1--12},
}

@article{almazrooie_quantum_2018,
	title = {Quantum reversible circuit of {AES}-128},
	volume = {17},
	issn = {1573-1332},
	url = {https://doi.org/10.1007/s11128-018-1864-3},
	doi = {10.1007/s11128-018-1864-3},
	language = {en},
	number = {5},
	urldate = {2025-09-03},
	journal = {Quantum Information Processing},
	author = {Almazrooie, Mishal and Samsudin, Azman and Abdullah, Rosni and Mutter, Kussay N.},
	month = mar,
	year = {2018},
	pages = {112},
}

@inproceedings{durmuth_towards_2021,
	address = {Cham},
	title = {Towards {Quantum} {Large}-{Scale} {Password} {Guessing} on {Real}-{World} {Distributions}},
	isbn = {978-3-030-92548-2},
	doi = {10.1007/978-3-030-92548-2_22},
	language = {en},
	booktitle = {Cryptology and {Network} {Security}},
	publisher = {Springer International Publishing},
	author = {Dürmuth, Markus and Golla, Maximilian and Markert, Philipp and May, Alexander and Schlieper, Lars},
	editor = {Conti, Mauro and Stevens, Marc and Krenn, Stephan},
	year = {2021},
	pages = {412--431},
}

@inproceedings{hossain_faruk_review_2022,
	title = {A {Review} of {Quantum} {Cybersecurity}: {Threats}, {Risks} and {Opportunities}},
	shorttitle = {A {Review} of {Quantum} {Cybersecurity}},
	url = {https://ieeexplore.ieee.org/document/9896970},
	doi = {10.1109/ICAIC53980.2022.9896970},
	urldate = {2025-09-03},
	booktitle = {2022 1st {International} {Conference} on {AI} in {Cybersecurity} ({ICAIC})},
	author = {Hossain Faruk, Md Jobair and Tahora, Sharaban and Tasnim, Masrura and Shahriar, Hossain and Sakib, Nazmus},
	month = may,
	year = {2022},
	pages = {1--8},
}

@article{zheng_implementation_2007,
	title = {Implementation of the {Grover} search algorithm with {Josephson} charge qubits},
	volume = {453},
	issn = {0921-4534},
	url = {https://www.sciencedirect.com/science/article/pii/S0921453406008756},
	doi = {10.1016/j.physc.2006.12.015},
	number = {1},
	urldate = {2025-09-03},
	journal = {Physica C: Superconductivity},
	author = {Zheng, Xiao-Hu and Dong, Ping and Xue, Zheng-Yuan and Cao, Zhuo-Liang},
	month = mar,
	year = {2007},
	pages = {76--79},
}
\end{document}